\documentclass{PoS}

\usepackage{amsmath}
\usepackage{graphicx}

\usepackage{amssymb}
\usepackage{bm}
\usepackage{cite}

\usepackage[normalem]{ulem}  

\renewcommand\sout{\bgroup \color{red} \ULdepth=-.5ex \ULset}

\newcommand{\be}{\begin{equation}}
\newcommand{\ee}{\end{equation}}
\newcommand{\ba}{\begin{align}}
\newcommand{\ea}{\end{align}}

\newcommand{\chibar}{{\bar{\chi}}}

\newcommand{\Fint}{{\cal D}}
\newcommand{\Seff}[2]{S_\mathrm{#1}^{(#2)}}

\newcommand{\Tr}{\mathrm{Tr}}

\def\Vec#1{\bm{#1}}

\newcommand{\mrm}[1]{\mathrm{#1}}
\newcommand{\Nt}{N_{\tau}}
\newcommand{\equref}[1]{Eq.~(\ref{#1})}

\newcommand{\figref}[1]{Fig.~\ref{#1}}

\title{
Fluctuation effects on QCD phase diagram at strong coupling
\footnote{Report No. : 
  YITP-14-74, 
  KUNS-2522
}
}

\ShortTitle{Fluctuation effects on QCD phase diagram at strong coupling}

\author{\speaker{Terukazu Ichihara}\\
          Department of Physics \&
	  Yukawa Institute for Theoretical Physics, \\
	  Kyoto University, \
          Kyoto 606-8502, Japan\\
	  E-mail: \email{t-ichi@ruby.scphys.kyoto-u.ac.jp}
}

\author{Akira Ohnishi\\
	    Yukawa Institute for Theoretical Physics, \\
	    Kyoto University,
             Kyoto 606-8502, Japan\\
	    E-mail: \email{ohnishi@yukawa.kyoto-u.ac.jp}
}

\abstract{
We study the QCD phase diagram away from the strong coupling limit (SCL)
with fluctuation effects
in the auxiliary field Monte-Carlo (AFMC) method.
First, we give an effective action which contains 
next-to-leading order (NLO) finite coupling effects 
of the strong coupling expansion 
as well as fluctuation effects.
Second, we examine NLO effects 
of the strong coupling expansion in AFMC at zero quark density.
We find that the chiral condensate is reduced by both
NLO terms from temporal plaquettes and fluctuation effects, 
and almost no dependence on NLO terms 
from spatial plaquettes in the current analysis.
These behaviors can be understood 
from the modification of the mass and the wave function renormalization factor
by auxiliary fields
as
in the mean field analysis 
and the fluctuation effects in the strong coupling limit.
}

\FullConference{The 32nd International Symposium on Lattice Field Theory\\
                 23-28 June, 2014\\
                 Columbia University New York, NY}

\begin{document}


\section{Introduction}
\label{sec:Intro}
Quantum Chromodynamics (QCD) phase diagram has been studied 
extensively, 
since it is related to many fields in physics ---
the early universe, heavy ion collisions, neutron stars and so forth.
However, the structure and phase boundary of the QCD phase diagram
is under debate mainly because of the notorious sign problem 
at finite chemical potential ($\mu$)
in the lattice QCD Monte-Carlo simulations.
There is a possibility that strong coupling lattice QCD (SC-LQCD) 
has a milder sign problem than the standard lattice QCD simulations, 
since the effective action can be written in terms of 
hadronic degrees of freedom.
In SC-LQCD, the sign problem does not exist in the mean field (MF) analysis,
and we could study the chiral phase transition \cite{SCL,large_d,Jolicoeur,NLOchiral,NLOPD,NNLO,Faldt,Bilic,MF-Pol}. 
When we include fluctuation effects, 
we generally have the sign problem 
as reported in monomer-dimer-polymer (MDP) simulation \cite{KarschMutter,SC-Rewei,MDP}
and auxiliary filed Monte-Carlo (AFMC) method \cite{AFMC}.
The sign problem in the strong coupling limit (SCL)
is not severe in these methods,
and we could investigate the QCD phase diagram 
even in the finite $\mu$ region.

SCL is the opposite limit of the continuum limit,
so we need to include finite coupling effects 
to get insight into the continuum QCD phase diagram.
In the MF analyses,
we find that finite coupling \cite{NLOPD,NNLO} and Polyakov loop effects 
reduce critical temperature at $\mu=0$ \cite{MF-Pol}, 
and the critical point (the endpoint of the first order phase transition) 
temperature also decreases with increasing inverse coupling.
On the other hand, the critical point chemical potential
evolves in a different way
in the cases where we include finite coupling effects
of the next-to-leading order 
(NLO, $\mathcal{O}(1/g^2)$) \cite{NLOPD}
and of the next-to-next-to-leading order 
(NNLO, $\mathcal{O}(1/g^4)$) \cite{NNLO}
in the MF treatment.
It is necessary to include both finite coupling and fluctuation effects 
to study the QCD phase diagram evolution approaching to the continuum region.
Both of these effects are evaluated
by using reweighing method in MDP \cite{SC-Rewei}. 
Thus
it is an important step to study the QCD phase diagram 
from directly sampled Monte-Carlo configurations.
In this paper, we construct an effective action including
finite coupling effects of NLO  
in AFMC, and
evaluate the fluctuation and finite coupling effects 
from the directly sampled configurations.

\section{NLO SC-LQCD effective action in Auxiliary field Monte-Carlo method}
\label{sec:formulation}
In the following analysis,
we only consider the case where the spatial lattice spacing $a$ is unity ($a=1$), and
the color is 3 ($\mathrm{SU}(N_c=3)$)
in 3+1 dimension $(d=3)$ spacetime.
Temporal and spatial lattice sizes are indicated as $N_\tau$ and $L$,
respectively.
The lattice QCD partition function and action
with one species of unrooted staggered fermion for color $\mathit{SU}(N_c)$ 
in the anisotropic Euclidean spacetime
are given by
\begin{align}
  {\cal Z}_{\mathrm{LQCD}} 
  =&  \int \Fint \left[ \chi,\chibar,U_\nu \right] e^{-S_\mathrm{LQCD}}
  \ , \ 
  S_\mathrm{LQCD}=S_F+S_G
  \ ,\\
  S_F
  =&\frac12 \sum_x \left[
   	V^{+}_x - V^{-}_x
  		\right]
  +\frac{1}{2\gamma} \sum_{x}\sum_{j=1}^{d} \eta_{j,x}\left[
  		 \bar{\chi}_x U_{j,x} \chi_{x+\hat{j}}
  		-\bar{\chi}_{x+\hat{j}} U^\dagger_{j,x} \chi_{x}
  		\right]
  +\frac{m_0}{\gamma} \sum_{x} M_x
\label{Eq:LQCD}
\ ,\\
V^{+}_x=& e^{\mu/f(\gamma)} \bar{\chi}_x U_{0,x} \chi_{x+\hat{0}}
\ ,
V^{-}_x= e^{-\mu/f(\gamma)} \bar{\chi}_{x+\hat{0}} U^\dagger_{0,x} \chi_x
\ ,
M_x=\bar{\chi}_x \chi_x
\ ,\\
S_G
=&
\frac {2N_c \xi}{g_{\mrm{\tau}}^2(g_0,\xi)} \mathcal{P}_\tau 
+ \frac {2N_c}{g_{\mrm{s}}^2(g_0,\xi) \xi} \mathcal{P}_s 
\ ,
\mathcal{P}_i
=
  \sum_{P_i} \left[
 	1 - \displaystyle \frac{1}{2N_c}  \Tr \left( U_{P_i} + U_{P_i}^\dagger \right) \right] 
\ (i=\tau,s)
\ ,
\end{align}
where
$\chi_x$, $U_{\nu,x}$, $U_{P_\tau}$ and $U_{P_s}$
represent the quark field, the link variable, 
and the temporal and spatial plaquettes,
respectively.
The quark chemical potential $\mu$ is introduced
in the form of the temporal component of vector potential, 
$\eta_{j,x}=(-1)^{x_0+\cdots+x_{j-1}}$ is
the staggered sign factor,  
and $V^{\pm}_x$ and $M_x$ are mesonic composites.
The temporal and spatial physical lattice spacing ratio is introduced as
$f(\gamma)=a/a_\tau$.
For simplicity, we here assume that the anisotropy parameters 
in the quark and gluon sectors
 are the same ($\gamma = \xi$),
and the temporal and spatial coupling constants are also
the same ($g_\tau = g_s =g$).
Under these assumptions,
the anisotropy is found to scale
the lattice spacing ratio as $f(\gamma)=\gamma^2$ and we can
define temperature as $T=\gamma^2/N_{\tau}$
from the mean field arguments
in SCL~\cite{Bilic}.
We utilize this prescription in the present work.
The above lattice QCD action 
has chiral symmetry $U(1)_V \times U(1)_A$
in the chiral limit $m_0\to 0$ . 
We obtain an NLO effective action 
after the spatial link integral as follows \cite{Jolicoeur,NLOPD,NNLO,Bilic,NLOchiral,Faldt}.
\begin{align}
\Seff{eff}{\mrm{NLO}} &= \Seff{eff}{\mrm{SCL}} + \Delta \Seff{}{\tau} + \Delta \Seff{}{s} + \mathcal{O}(1/\sqrt{d},1/g^4)
\ ,\  \\
 \Seff{eff}{\mrm{SCL}} 
&=\frac12 \sum_x \left[ V^{+}_x - V^{-}_x \right]
- \displaystyle \frac {1}{4N_c\gamma} \sum_{x,j} M_x M_{x+\hat{j}}
+\frac{m_0}{\gamma} \sum_{x} M_x
\ , \\
\Delta \Seff{}{\tau} &=\frac{1}{2}C_\tau \sum_{x,j>0} \left[ V_{x}^{+}(\mu) V_{x+\hat{j}}^{-} (\mu) + V_{x}^{+}(\mu) V_{x-\hat{j}}^{-} (\mu) \right]
\ , \\
\Delta \Seff{}{s} &= -\frac{1}{4N_{c} \gamma^4 }C_s \sum_{x,0<k<j} M_{x}M_{x+\hat{j}}M_{x+\hat{k}}M_{x+\hat{j}+\hat{k}}
\ , 
\end{align}
where $C_\tau =1/(2N_c^2 g^2 \gamma)$ and $C_s= 1/ (2 N_c^3 g^2 \gamma)$.
We only consider the leading order terms of the large diminutional ($1/d$) expansion \cite{large_d}.

We now apply the 
extended Hubaard-Stratonovich transformaion 
~\cite{AFMC} in order 
to reduce the interaction terms of quarks into the bilinear form.
As for the spatial NLO terms, we utilize sequential
bosonization.
First, the bosonized effective action for spatial NLO terms is given as
\begin{align}
\Seff{sNLO0}{\mrm{EHS}}
&= \frac{L^3 C_{s}}{8N_{c}} 
 \sum_{\tau,\Vec{u},\kappa_u^j >0,j} \kappa_{u}^{(j)} \left[ 
   | \Sigma_{\Vec{u},\tau}^{(j)} |^2  +| \Pi_{\Vec{u},\tau}^{(j)} |^2
   \right] + \frac{C_s}{4N_{c}\gamma^2} 
 \sum_{x,j} C_{j,x}^s M_{x}M_{x+\hat{j}}
 \ , \ 
 \label{Eq:SeffsNLO0}
\end{align}
where $\kappa _u^{(\mrm{j})} = \sum_{k(\neq j)}\cos u_k$, $\Sigma_{x}^{(j)} = \sum_{\Vec{k},\kappa_u^j>0} e^{i\Vec{k}\cdot\Vec{x}} \tilde{\Sigma}^{(j)}_{\Vec{k}}(\tau) $ , $\Pi_{x}^{(j)}=(-1)^{\tau}\sum_{\Vec{k},\kappa_u^j>0} e^{i\Vec{k}\cdot\Vec{x}} \tilde{\Pi}^{(j)}_{\Vec{k}}(\tau) $, 
$C_{j,x}^{s} = \sum_{k(\neq j)} \left[  (\Sigma^{(j)} + i \epsilon \Pi^{(j)})_{x+\hat{k}} +
(\Sigma^{(j)} + i \epsilon \Pi^{(j)})_{x-\hat{k}} \right]$,
and $\epsilon_x=(-1)^{x_0+\cdots +x_d}$ which is related to $\gamma_5$ in the continuum theory.
In \equref{Eq:SeffsNLO0}, we divide  the momentum region 
according to positive $(\kappa_{\bm{u}}^{(j)}>0)$ and negative $(\kappa_{\bm{u}}^{(j)}<0)$ modes, and use the relation $\kappa_{\bar{\bm{u}}}^{(j)} = -\kappa_{\bm{u}}^{(j)}$, where $\bar{\bm{u}}=\bm{u}+(\pi,\pi,\pi)$ as SCL~\cite{AFMC}. 
Second, we use the EHS transformation to the spatial NLO action, $\Seff{sNLO0}{\mrm{EHS}}$, once again.
\begin{align}
\Seff{sNLO}{\mrm{EHS}} =  \frac{C_s}{4N_c} \sum_{x} \left[ \phi_{x}^2 +\varphi_x^2 + i \left\{  \left( \frac{M_x}{\gamma}  + \frac{\sum_j C_{j,x}^s M_{x+\hat{j}} }{\gamma}  \right) \varphi_{x} -i  \left( \frac{M_x}{\gamma}  - \frac{\sum_j C_{j,x}^s M_{x+\hat{j} } }{\gamma} \right) \phi_x \right\}  \right]
 \ .
\label{Eq:SeffsNLO}
\end{align}
%
We also obtain the bosonized action terms of
temporal NLO terms as
\begin{align}
\Seff{tNLO}{\mrm{EHS}} &= 
L^3 C_\tau \sum_{\tau,\bm{k},f(\bm{k})>0}f(\bm{k}) \left[ |\omega_{\bm{k},\tau}|^2+|\Omega_{\bm{k},\tau}|^2 \right]  + \frac{1}{2}\sum_{x} \left[ V_x^+ (\mu) \Delta \alpha_x^-   - V_x^- (\mu) \Delta \alpha_x^+ \right]
\ ,
\label{Eq:SefftNLO}
\end{align}
\begin{align}
\Delta \alpha_x^- 
&= C_{\tau}\sum_j \left\{-i \omega_{x+\hat{j}}^* -i\omega_{x-\hat{j}}^* - (\epsilon \Omega^* )_{x+\hat{j}} -(\epsilon \Omega^*)_{x-\hat{j}} \right\}
\ ,
\\
\Delta \alpha_x^+  
&
= C_{\tau} \sum_j \left\{
	i \omega_{x+\hat{j}} + i\omega_{x-\hat{j}} 
	+ (\epsilon \Omega)_{x+\hat{j}}  + (\epsilon \Omega)_{x-\hat{j}}
\right\}
\ ,
\end{align}
where 
$f(\bm{k})=\sum_{j} \cos k_j$.
 $\omega_{x} = 
  \sum_{\Vec{k},f(\Vec{k})>0} e^{i\Vec{k}\cdot\Vec{x}} \tilde{\omega}_{\Vec{k}}(\tau) $, 
$  \Omega_{x}=  (-1)^{\tau}\sum_{\Vec{k},f(\Vec{k})>0} e^{i\Vec{k}\cdot\Vec{x}} \tilde{\Omega}_{\Vec{k}}(\tau) $. 
With
the SCL terms~\cite{AFMC},
the total effective action after EHS is given as
\begin{align}
\Seff{eff}{\mrm{EHS}}  &=
 \frac{L^3 C_{s}}{8N_{c}} 
 \sum_{\tau,\Vec{u},\kappa_u^j >0, j} \kappa_{u}^{(j)} \left[ 
   | \Sigma_{\Vec{u},\tau}^{(j)} |^2  +| \Pi_{\Vec{u},\tau}^{(j)} |^2
   \right]
 +L^3 C_\tau \sum_{\tau,\bm{k},f(\bm{k})>0}f(\bm{k}) \left[ |\omega_{\bm{k},\tau}|^2+|\Omega_{\bm{k},\tau}|^2 \right] 
\nonumber \\
&
 + \frac{L^3}{4N_c} \sum_{\bm{k},\tau,f(\bm{k})>0} f(\bm{k}) \left[ |\sigma_{\bm{k},\tau}|^2 +|\pi_{\bm{k},\tau}|^2 \right]
 + \frac{C_s}{4N_c} \sum_{x}\left[ \phi_x^2 +\varphi_x^2 \right]
\nonumber \\
&+ \frac{1}{2}\sum_{x} Z_{x}\left[ V_{x}^{+}(\tilde{\mu}) -  V_{x}^{-}(\tilde{\mu})  \right]
  +\sum_x m_x M_x / \gamma
\  , 
\label{Eq:SeffEHS}
\end{align}
where
$Z_{x}=\sqrt{\alpha^{+}_{x}\alpha^{-}_{x}}\ , 
\ e^{\tilde{\mu}_x}=e^{\mu} e^{-\delta \mu _x} =e^{\mu} \sqrt{\alpha^{-}_{x}/\alpha^{+}_{x}} \ ,
\alpha_x^+ = 1+ \Delta \alpha_x^+\ ,\ \alpha_x^- = 1+ \Delta \alpha_x^-$, 
\begin{align}
  m_x &= m_0 +\frac{1}{4N_c } \sum_j 
  \left[ (\sigma + i \epsilon \pi)_{x-\hat{j}} +(\sigma + i\epsilon \pi)_{x+\hat{j}} \right]
    +\frac{C_s}{4N_c}  i \left[ \left( \varphi_x -i\phi_x \right) 
    + \sum_j \left( C_{j,x-\hat{j}}^s \varphi_{x-\hat{j}} 
         + i C_{j,x-\hat{j}}^s \phi_{x-\hat{j}} \right) \right]
\  .
\end{align}
It is clear that the mass and chemical potential are modified 
by the auxiliary fields.
In the temporal kinetic terms, the wave function renormalization factor $Z_x$
appears as in the MF analysis \cite{NLOPD}.

Next, we integrate out Grassmann variables and temporal link variables analytically, then the effective action is written in terms of hadronic degrees of freedom as
\begin{align}
\Seff{eff}{\mrm{AF}} &=  \frac{L^3 C_{s}}{8N_{c}} 
 \sum_{\tau,\Vec{u},\kappa_u^j >0,j} \kappa_{u}^{(j)} \left[ 
   | \Sigma_{\Vec{u}}^{(j)} |^2  +| \Pi_{\Vec{u}}^{(j)} |^2
   \right]
 +L^3 C_\tau \sum_{\tau,\bm{k},f(\bm{k})>0} f(\bm{k}) \left[ |\omega_{\bm{k},\tau}|^2+|\Omega_{\bm{k},\tau}|^2 \right] 
\nonumber \\
&
 + \frac{L^3}{4N_c} \sum_{\bm{k},\tau,f(\bm{k})>0} f(\bm{k}) \left[ |\sigma_{\bm{k},\tau}|^2 +|\pi_{\bm{k},\tau}|^2 \right]
 + \frac{C_s}{4N_c} \sum_{x}\left[ \phi_x^2 +\varphi_x^2 \right]
\nonumber \\
&  -  \sum_{\bm{x}}
  \log \left[
  X_{\Nt}(\bm{x})^3 -2 \hat{Z}(\bm{x})^2 X_{\Nt} (\bm{x})+ \hat{Z}(\bm{x})^3 2\cosh \left( 3 \hat{\tilde{\mu}} (\bm{x}) \right)
  \right]
  \ . 
\label{Eq:SeffNLO}
\end{align}
where $\hat{Z} (\bm{x}) = \prod_{i} Z_{\bm{x},i}$ and 
$\hat{\tilde{\mu}} (\bm{x}) = \prod_{i} \tilde{\mu}_{\bm{x},i}
 \ $.
We evaluate $X_{N_\tau}$ using the recursion 
formulae~\cite{Faldt}.
In the last step, we integrate out auxiliary fields
by Monte-Carlo technique at a time. We generate Monte-Carlo configurations and investigate auxiliary field fluctuation effects in AFMC.

\section{Chiral condensate in NLO with fluctuations}

In order to 
understand the finite coupling effects 
from temporal and spatial plaquettes separately,
we examine the phase transition at $\mu=0$ with two effective actions:
 one is the action in which 
we consider
 temporal NLO 
and SCL
auxiliary fields
named as t-NLO,
and the other is the action
in which we take account of 
 spatial NLO and SCL terms
named as sp-NLO.

\begin{figure}[tb]
 \begin{center}
   \includegraphics[width=55mm,angle=270]{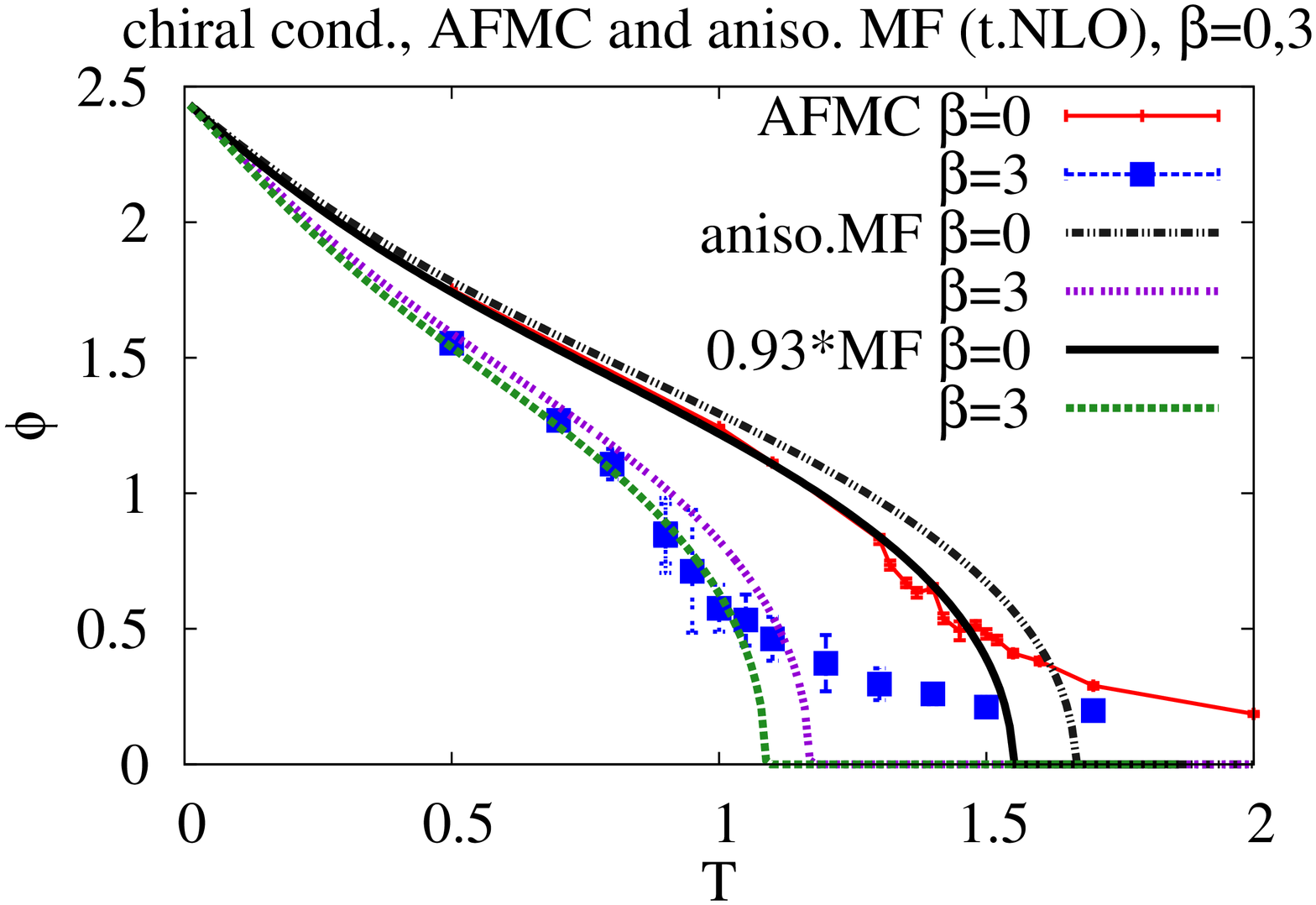}
   \includegraphics[width=55mm,angle=270]{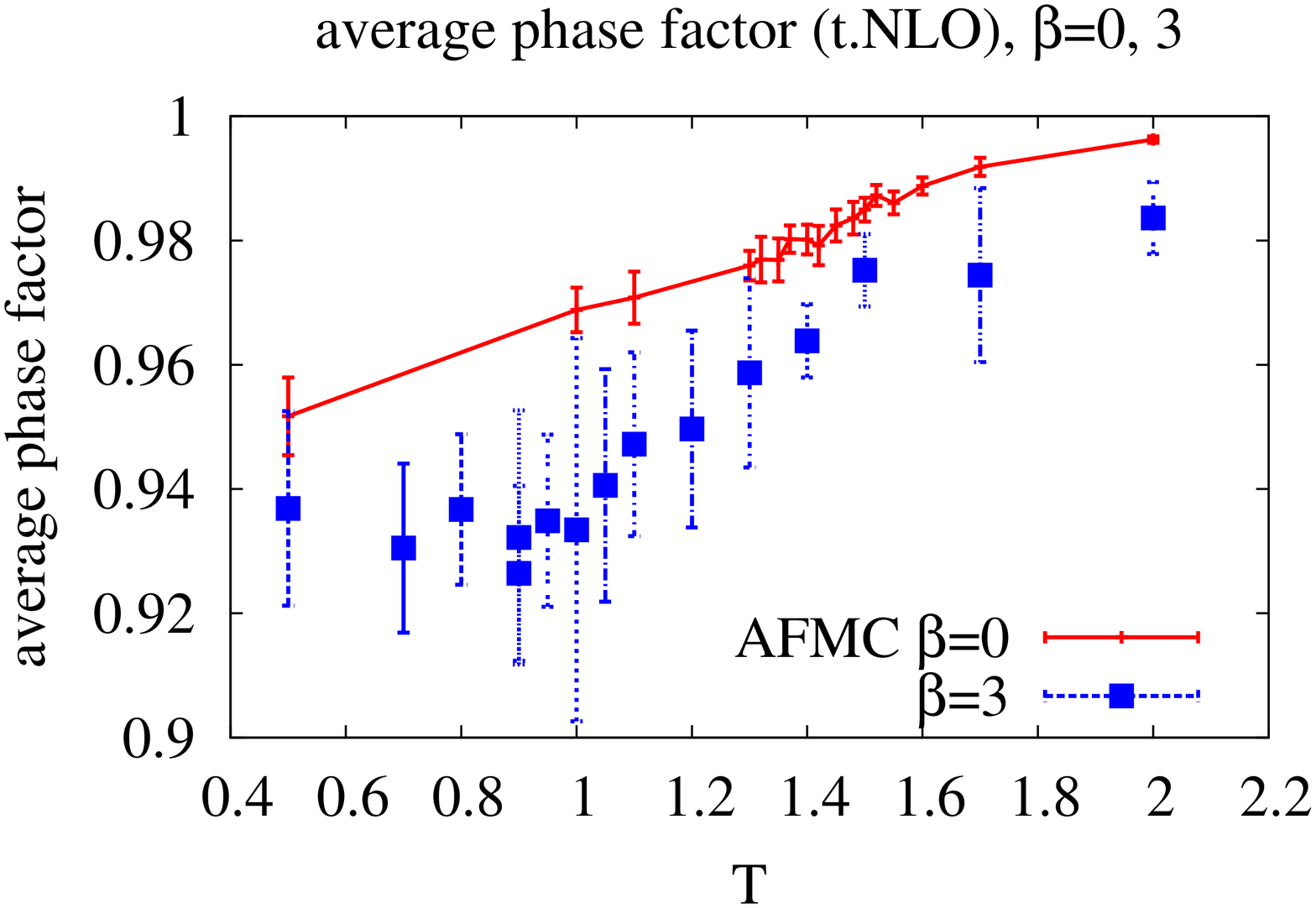}\\
 \includegraphics[width=53mm,angle=270]{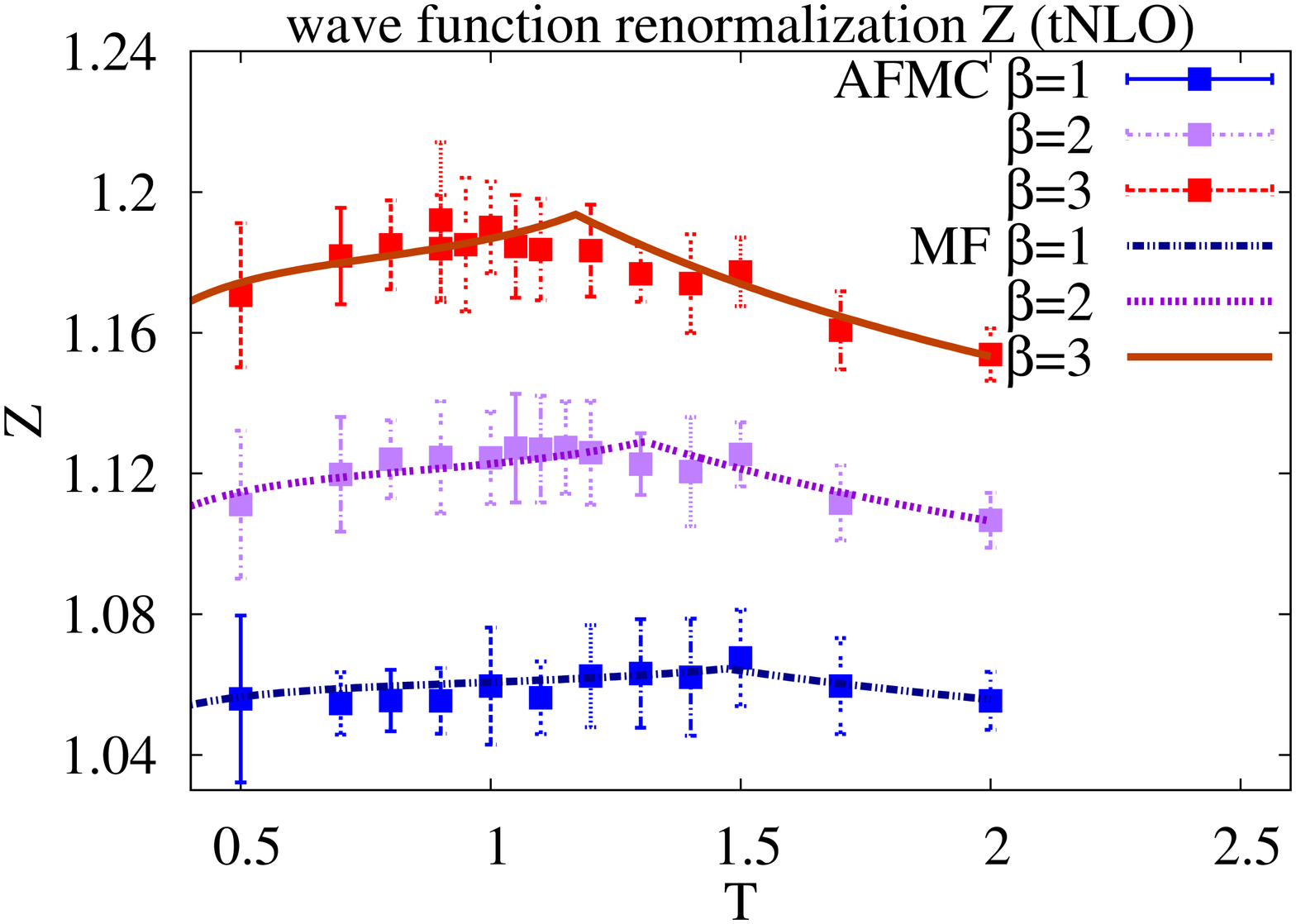}
   \includegraphics[width=55mm,angle=270]{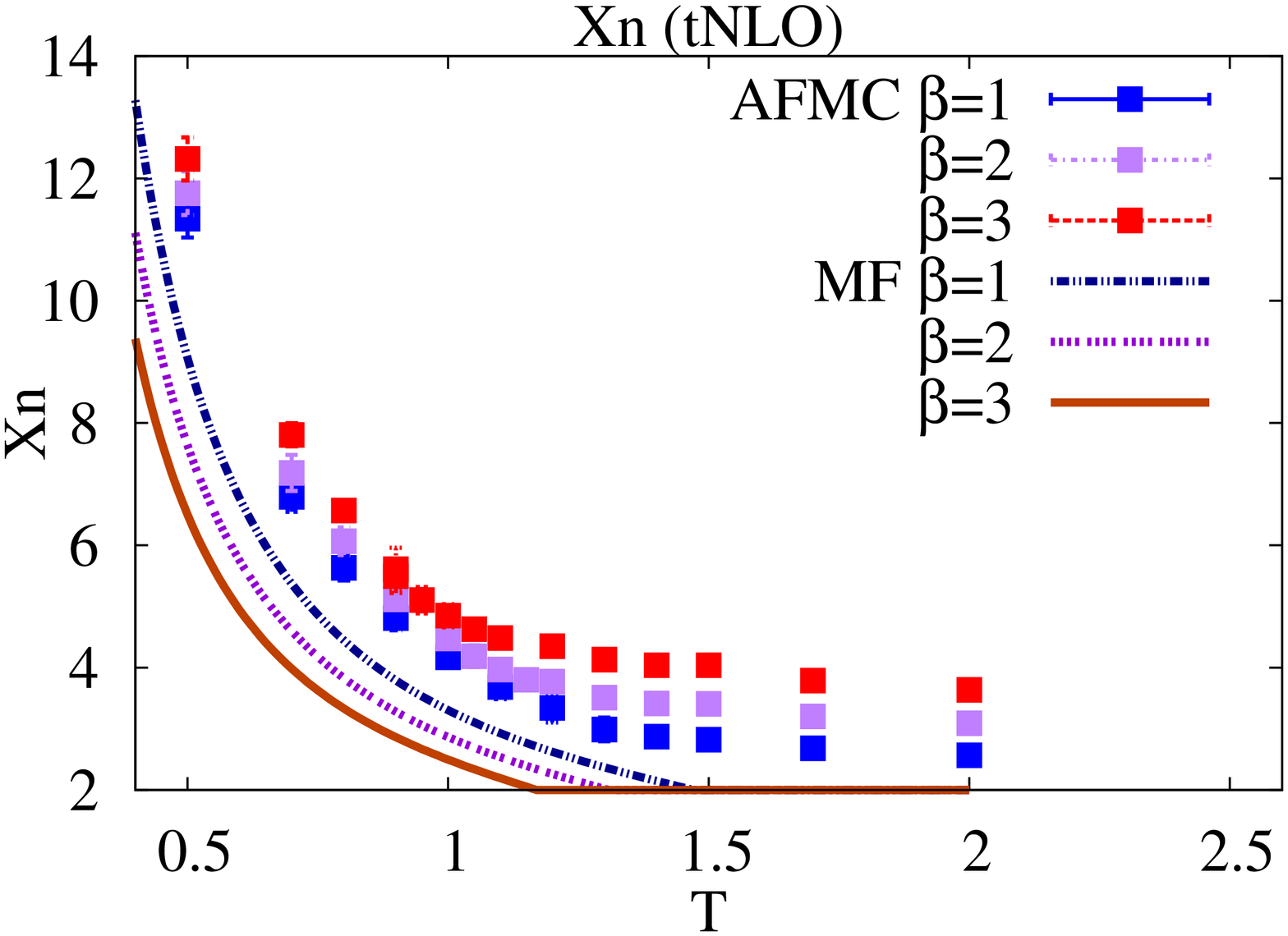}
  \end{center}
\caption{The chiral condensate (top left), 
average phase factor (top right),
wave function renormalization $Z$ (bottom left), 
and $X_{N_\tau}$ (bottom right) 
as functions of temperature at $\mu=0$
for $\beta=2N_c /g^2 =0,\ 1,\ 3$ in t-NLO.
The chiral condensate in AFMC is compared with the MF results,
and also with the MF results with reduced temperature, $0.93 \times T$.
}
 \label{Fig:tNLOsig}
\end{figure}
First, we show the chiral condensate in t-NLO
at $\mu=0$
on a $4^3 \times 4$ lattice in \figref{Fig:tNLOsig}.
We here regard the chiral radius as the chiral condensate.
We find that the chiral condensate is reduced by both fluctuation and finite coupling effects.
The main finite coupling effect is the mass reduction
via the wave function renormalization factor $Z_{x}$
as discussed in MF analyses~\cite{NLOPD}.
Since $Z_x$ coming from the temporal auxiliary fields
is larger than unity as shown in \figref{Fig:tNLOsig},
the effective mass $m_x /Z_x$ is 
suppressed, 
then the chiral condensate is 
reduced~\cite{NLOPD}.
By comparison, fluctuations modify $X_{\Nt}$ and play a role to enhance
the apparent temperature:
the chiral condensate in AFMC agrees well with the MF result
at enhanced temperature,
$\phi_\mathrm{AFMC}(T) \simeq \phi_\mathrm{MF}(T/0.93)$.
To our surprise,
fluctuation effects and finite coupling effects suppress 
the chiral condensate cumulatively
since the both results at zero and finite $\beta$ in AFMC are well described by 93 percent scaling results in MF.
In t-NLO simulation, 
the average phase factor is larger than 0.9 and the wight cancellation is not severe.
Using the MF knowledge, 
the repulsive force coming from the auxiliary field, $\omega_t$, leads to severe sign problem.
However, the field $\omega_t$ vanishes at zero $\mu$ and do not affect the sign problem in MF.
We could expect that the average phase factor is large enough to simulate observables even in AFMC at zero $\mu$.

\begin{figure}[tbh]
\begin{minipage}{0.5\hsize}
 \begin{center}
   \includegraphics[width=55mm,angle=270]{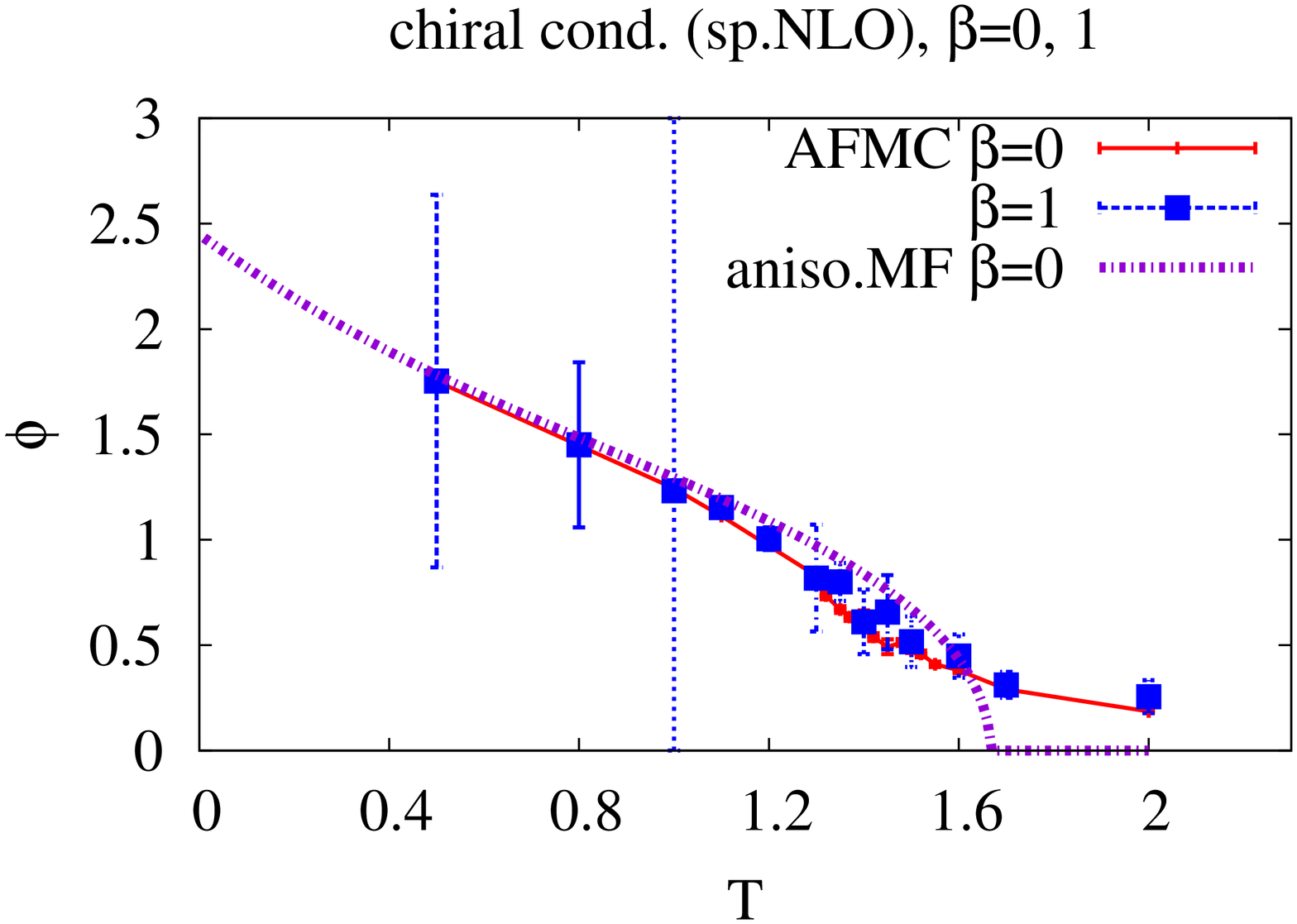}%
  \end{center}
  \end{minipage}
  \begin{minipage}{0.5\hsize}
  \begin{center}
   \includegraphics[width=55mm,angle=270]{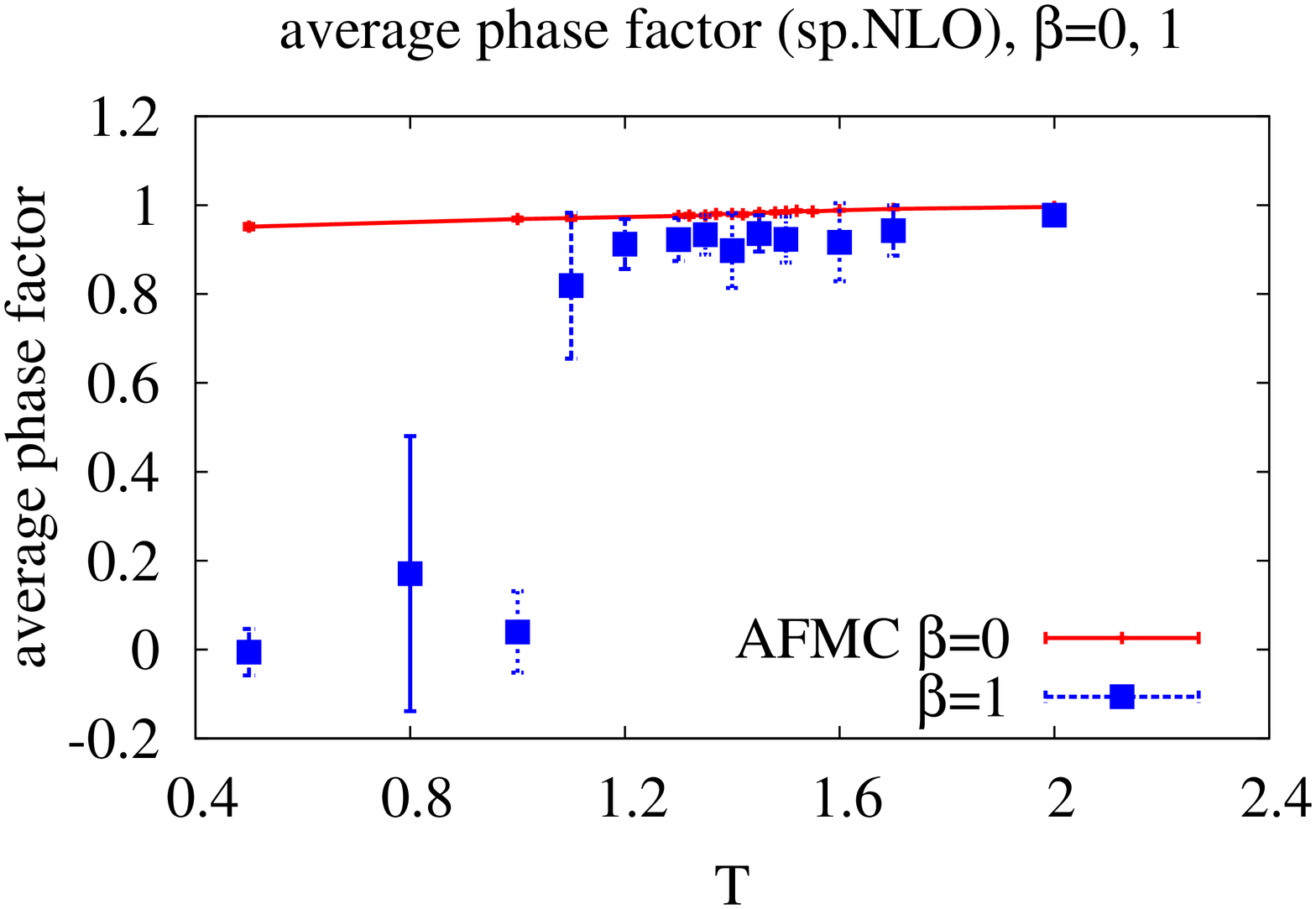}%
  \end{center}
  \end{minipage}
 \caption{The chiral condensate and average phase factor as a function of temperature for $\beta = 0,1$ with sp. NLO.
  The chiral condensate is almost the same value up to current analysis.
  Average phase factor is almost zero at low temperature, which means the sign problem is so severe.}
 \label{Fig:sNLOsig}
\end{figure}

In \figref{Fig:sNLOsig}, we show the chiral condensate and average phase factor in sp-NLO.
The chiral condensate takes almost the same value 
as MF results,
while the average phase factor is smaller than that in t-NLO 
and is consistent with zero at low temperatures.
We need to develop some ways to weaken the sign problem in AFMC with spatial NLO terms.

\section{Summary}
We have studied the fluctuation and finite coupling effects
on QCD phase diagram in the framework of the strong coupling expansion
and auxiliary fields Monte-Carlo (AFMC) method.
We have constructed an effective action of auxiliary fields with both fluctuation and finite coupling effects.
We apply the Monte-Carlo technique to integrate over auxiliary fields 
 in order to take fluctuation effects into account.
We have numerically examined the finite coupling effects from temporal
and spatial plaquettes separately.
When
we take account of strong coupling limit (SCL) and temporal NLO auxiliary fields denoted as t-NLO, both of finite coupling and fluctuation effects are found to reduce the chiral condensate.
In t-NLO simulations,  we find that the average phase factor is large enough at zero chemical potential.
When we consider SCL and spatial NLO auxiliary fields (sp-NLO),
the chiral condensate are not 
modified much in the current analysis.
The average phase factor is 
small, and it is necessary to develop
a new way to weaken or avoid the sign problem in AFMC 
in order to obtain results in sp-NLO and full NLO simulation.

\section*{Acknowledgments}
The authors would like to thank Wolfgang Unger, Frithjof Karsch, Swagato Mukherjee 
and participants of 
the 32nd Int. Symp. on Lattice Field Theory (Lattice 2014)
for useful discussions.
TI is supported by the Grants-in-Aid for JSPS Fellows (No.25-2059). 
This work is supported in part by the Grants-in-Aid for Scientific Research
from JSPS
(Nos.
           23340067, 
           24340054, 
           24540271
),
by the Grants-in-Aid for Scientific Research on Innovative Areas from MEXT
(No. 2404: 24105001, 24105008), 
by the Yukawa International Program for Quark-hadron Sciences,
and by the Grant-in-Aid for the global COE program ``The Next Generation
of Physics, Spun from Universality and Emergence" from MEXT.



\begin{thebibliography}{99}
\bibitem{SCL}
  N.~Kawamoto and J.~Smit, Nucl.~Phys.~{\bf B192}, 100 (1981); 
  H.~Kluberg-Stern, A.~Morel, B.~Petersson, 
  Phys.\ Lett.\ B {\bf 114}, 152 (1982);
  J.~Hoek, N.~Kawamoto, and J.~Smit,
  Nucl.~Phys.~B {\bf 199}, 495 (1982); 
  P.~H.~Damgaard, N.~Kawamoto and K.~Shigemoto,
  Phys.\ Rev.\ Lett.\  {\bf 53}, 2211 (1984); 
 P.~H.~Damgaard, D.~Hochberg, and N.~Kawamoto, 
  Phys.~Lett.~ B {\bf 158}, 239 (1985);
 P.~H.~Damgaard, N.~Kawamoto, and K.~Shigemoto, 
  Nucl.~Phys.~B {\bf 264}, 1 (1986); 
 V.~Azcoiti, G.~Di Carlo, A.~Galante, and V.~Laliena, 
 J.~High Energy Phys.~{\bf 09}, 014 (2003);
  K.~Fukushima,
  Prog.\ Theor.\ Phys.\ Suppl.\  {\bf 153}, 204 (2004)
  [hep-ph/0312057];
  Y.~Nishida,
  Phys.\ Rev.\ D {\bf 69}, 094501 (2004)
  [hep-ph/0312371].

\bibitem{large_d}
  H.~Kluberg-Stern, A.~Morel, B.~Petersson, 
  Nucl.~Phys.~B {\bf 215} [FS7], 527 (1983);


\bibitem{Faldt}
  G.~Faldt and B.~Petersson,
  Nucl.\ Phys.\ B {\bf 265}, 197 (1986).

\bibitem{Bilic}
  N.~Bilic, K.~Demeterfi and B.~Petersson,
  Nucl.\ Phys.\ B {\bf 377}, 651 (1992);


  N.~Bilic, F.~Karsch, and K.~Redlich,
  Phys.~Rev.~D {\bf 45}, 3228 (1992); 
  N.~Bilic and J.~Cleymans,
  Phys.~Lett.~B {\bf 355}, 266 (1995).

\bibitem{Jolicoeur} 
  T.~Jolicoeur, H.~Kluberg-Stern, M.~Lev, A.~Morel, and B.~Petersson, 
  Nucl.~Phys.~B {\bf 235}, 455 (1984).

\bibitem{NLOchiral}
  I.~Ichinose, Phys.~Lett.~{\bf B135}, 148 (1984);
  ibid.~{\bf B147}, 449 (1984); 
  I.~Ichinose, Nucl.~Phys.~{\bf B249}, 715 (1985).
\bibitem{NLOPD}
  K.~Miura, T.~Z.~Nakano and A.~Ohnishi,
  Prog.\ Theor.\ Phys.\  {\bf 122}, 1045 (2009)
  [arXiv:0806.3357 [nucl-th]];
  K.~Miura, T.~Z.~Nakano, A.~Ohnishi and N.~Kawamoto,
  Phys.\ Rev.\ D {\bf 80}, 074034 (2009)
  [arXiv:0907.4245 [hep-lat]].

\bibitem{NNLO}
  T.~Z.~Nakano, K.~Miura and A.~Ohnishi,
  Prog.\ Theor.\ Phys.\  {\bf 123}, 825 (2010)
  [arXiv:0911.3453 [hep-lat]];
  T.~Z.~Nakano, K.~Miura and A.~Ohnishi,
  Phys.\ Rev.\ D {\bf 83}, 016014 (2011)
  [arXiv:1009.1518 [hep-lat]].

\bibitem{MF-Pol}
T.~Z.~Nakano, K.~Miura, and A.~Ohnishi,
Phys.~Rev.~D {\bf 83}, 016014 (2011);
T.~Z.~Nakano, K.~Miura, and A.~Ohnishi,
PoS LATTICE2010, 205 (2010);
K.~Miura, T.~Z.~Nakano, A.~Ohnishi, and N.~Kawamoto,
PoS LATTICE2011, 318 (2011) ; 
K.~Miura, T.~Z.~Nakano, A.~Ohnishi, and N.~Kawamoto,
arXiv:1106.1219 (2011).

\bibitem{KarschMutter}
  F.~Karsch and K.~H.~Mutter,
  Nucl.\ Phys.\ B {\bf 313}, 541 (1989).

\bibitem{MDP}
  P.~de Forcrand and M.~Fromm,
  Phys.\ Rev.\ Lett.\  {\bf 104}, 112005 (2010)
  [arXiv:0907.1915 [hep-lat]];
  W.~Unger and P.~de Forcrand,
  J.\ Phys.\ G {\bf 38}, 124190 (2011)
  [arXiv:1107.1553 [hep-lat]];
    M.~Fromm, 
  Ph.D. thesis, ETH-19297, Eidgen$\mathrm{\ddot{o}}$ssische Technische Hochschule ETH Z$\mathrm{\ddot{u}}$rich, 2010;
  M.~Fromm, P.~de Forcrand,
  PoS {\bf LATTICE2009}, 193 (2009). 


\bibitem{SC-Rewei}
  P.~de Forcrand, J.~Langelage, O.~Philipsen, and W.~Unger,
  PoS {\bf LATTICE2013}, 142 (2013) 
  [arXiv:1312.0589 [hep-lat]];
  W.~Unger,
  Acta Phys.~Polon.~Supp.~{\bf 7 No.~1}, 127 (2014);
  Ph.~de Forcrand, J.~Langelage, O.~Philipsen, and W.~Unger,
  Phys.~Rev.~Lett.~{\bf 113}, 152002 (2014). 
  


\bibitem{AFMC}
  A.~Ohnishi, T.~Ichihara and T.~Z.~Nakano, 
     PoS\ {\bf LATTICE2012}, 088 (2012), 
  T.~Ichihara, T.~Z.~Nakano, and A.~Ohnishi, 
     PoS\ {\bf LATTICE2013}, 143 (2013);
     T.~Ichihara, A.~Ohnishi, and T.~Z.~Nakano,
      arXiv:1401.4647.

\end{thebibliography}
\end{document}